\documentclass[aps,prl,showpacs,amsmath,twocolumn,amssymb,superscriptaddress,letterpaper]{revtex4}
\usepackage{graphicx,color}
\usepackage{amssymb}   
\usepackage{amsmath}
\usepackage{epstopdf}
\usepackage{natbib}
\usepackage{hyperref}
\usepackage{bm}
\begin{document}

\title{Flux-induced Topological Superconductor in Planar Josephson Junction}
\author{Jie Liu}
\affiliation{Department of Applied Physics, School of Science, Xi'an Jiaotong University, Xi'an 710049, China}
\author{Yijia Wu}
\affiliation{International Center for Quantum Materials, School of Physics, Peking University, Beijing 100871, China}
\author{Qing-Feng Sun}
\affiliation{International Center for Quantum Materials, School of Physics, Peking University, Beijing 100871, China}
\affiliation{Collaborative Innovation Center of Quantum Matter, Beijing 100871, China}
\affiliation{Beijing Academy of Quantum Information Sciences, Beijing 100193, China}
\author{X. C. Xie}
\affiliation{International Center for Quantum Materials, School of Physics, Peking University, Beijing 100871, China}
\affiliation{Beijing Academy of Quantum Information Sciences, Beijing 100193, China}
\affiliation{CAS Center for Excellence in Topological Quantum Computation,
University of Chinese Academy of Sciences, Beijing 100190, China}

\begin{abstract}
A planar Josephson junction with a normal metal attached on its top surface will form a hollow nanowire structure due to its three dimensional nature. In such hollow nanowire structure, the magnetic flux induced by a small magnetic field (about 0.01T) will tune the system into topologically non-trivial phase and therefore two Majorana zero-modes will form at the ends of the nanowire. Through tuning the chemical potential of the normal metal, the topologically non-trivial phase can be obtained for almost all energy within the band. Furthermore, the system can be conveniently tuned between the topologically trivial and non-trivial phases via the phase difference between the superconductors. Such device, manipulable through flux, can be conveniently fabricated into desired 2D networks. Finally, we also propose a cross-shaped junction realizing the braiding of Majorana zero-modes through manipulating the phase differences.
\end{abstract}

\maketitle

{\emph {Introduction}} --- Majorana zero-mode (MZM), the most promising candidate for topological quantum computation \cite{kitaev, nayak}, the search for its hiding place, topological superconductor (TSC) platforms has become one of the most exciting research areas in the past decade. Considerable theories \cite{Fu, sau, fujimoto, sato, alicea2, lut, oreg, potter, Ki2,Sup1} and experiments \cite{kou,deng,das1,hao1,hao2,Marcus,perge,Yaz2,Jia,Fes1, Fes2, Fes3} have contributed to the realization of TSC to date. A large number of systems, including semiconductor nanowires \cite{kou,deng,das1,hao1,hao2,Marcus}, ferromagnetic atomic chains \cite{perge, Yaz2}, and vortices on the surface of 3D topological insulators \cite{Jia}, will exhibit signals of MZMs in the presence of superconductivity proximity effect. Furthermore, evidences for MZM have also been reported for vortex in an intrinsic TSC based on iron-based superconductor \cite{Fes1, Fes2, Fes3}. However, the formation of MZMs at the ends of 1D nanowires or in the vortex cores
on 2D surface is only the first step towards the realization of topological
quantum computation. A promising candidate system for topological
quantum computation should possess following advantages simultaneously.
First of all, considering the braiding process of MZMs,
the system should be manipulable and could be easily tuned between topologically trivial
and non-trivial phases. In previous semiconductor nanowires,
the topology can only be tuned by gate voltages which generally
require a high level of sophistication \cite{alicea3,Jie1, Jie2}.
Secondly, the spatial scale of the system should be suitable to fabricate
circuits for quantum computation, therefore the vortex systems are obviously
not good choices.
Thirdly, a large and hard gap is also indispensable for the manipulation
of MZMs which usually requires a weak magnetic field \cite{har1}.
Therefore, though great progress has been made in the last decade,
the desired high quality TSC systems still remain to be explored.

\begin{figure}
\centering
\includegraphics[width=3.25in]{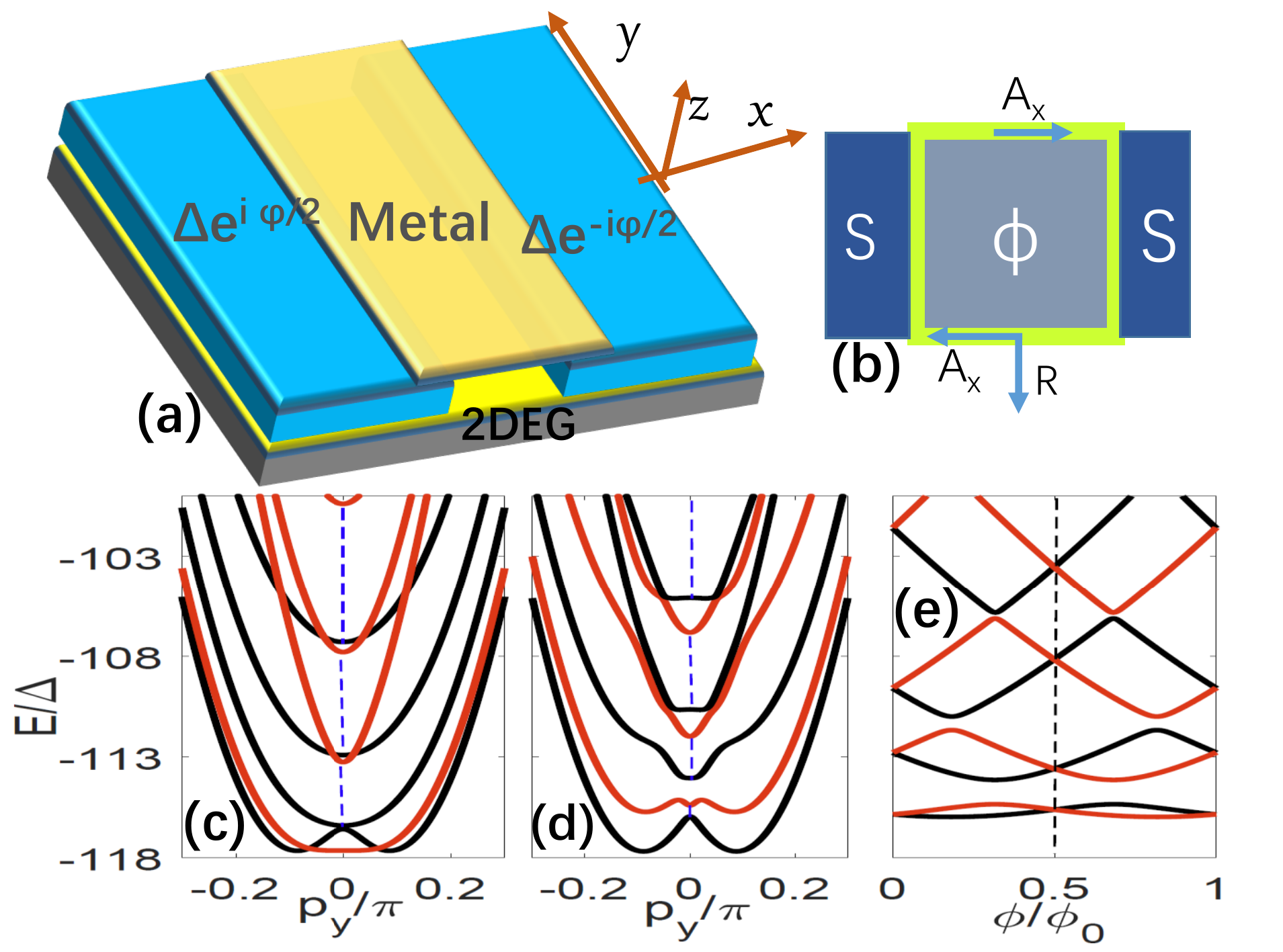}
\caption{
(a) A schematic setup for a normal metal attached on a planar Josephson junction
to form a superconductor-hollow nanowire-superconductor system.
(b) The device shown in (a) can be simplified into a hollow nanowire structure (cross section view).
(c) The energy spectrum of the rectangular hollow nanowire structure [shown in (b)]
with Rashba SOC along the radial direction at $\phi=0.25\phi_0$. The degeneracy of the bands is lifted by threading a flux.
(d) The energy spectrum of the rectangular hollow nanowire structure with Rashba SOC only presented at the bottom surface with $\phi =0.25\phi_0$.
(e) The band energy at $p_y=0$ versus the magnetic flux. The system is even occupied in the absence of flux, while the degeneracy is lifted with the increase of flux.
In (c)-(e), the sizes of the hollow nanowire are $W=5a$ and $H=5a$.
}
\label{f1}
\end{figure}

A recent progress points out that through patterning the TSC into a planar Josephson junction on a two dimensional electron gas (2DEG) \cite{2DEG1, 2DEG2},
we can conveniently fabricate a network structure and tune the topology
through the phase differences. In addition, by tuning the difference between superconducting phases, the required magnetic field can be reduced. Such structure almost meets all the requirements of topological quantum computation, therefore it has drawn great attention and has recently been experimentally realized \cite{PJJ1, PJJ2}. However, a relatively large magnetic field is still required as shown in the experiments. In the presence of large magnetic field, a small tilt of the magnetic field may destruct the band gap as well as the topological phase.
Furthermore, the topologically non-trivial region is only presented in a small range of energy at the bottom of the band, which is determined by the spin-orbit coupling energy \cite{2DEG2}.
For these reasons, an improved version for such planar Josephson junction
is highly needed.

In this paper, we propose a modified planar Josephson junction structure as depicted
in Fig. \ref{f1}(a). Through attaching a normal metal on the top surface of the planar Josephson junction, the system is confined to a hollow nanowire structure in the junction.
Then a weak magnetic field (smaller than 0.1T in principle) is threaded along the hollow nanowire,
where the magnetic flux can tune the system into TSC phase \cite{core1,core2}
and therefore two MZMs appear at the ends of the nanowire.
The normal metal also provides an additional free parameter that through tuning its chemical potential, the topologically non-trivial phase can be obtained for almost all the energy within the
band. Besides, the topologically non-trivial phase is found to be more stable in the band center than in the band bottom.
Such hollow nanowire also inherits the advantages of the previous planar Josephson junction.
Through manipulating the phase difference between the two sides of
the Josephson junction, the system can turn from topologically trivial
to topologically non-trivial phase and vice versa \cite{WPei}.
Such structure, tunable through flux,
can be easily fabricated into a 2D network.
In comparison, in the previously proposed planar Josephson junction
or semiconductor nanowire, the TSC can only be spatially extended along one direction.
Lastly, we also suggest a cross-shaped junction to realize the braiding of MZMs.

{\emph {Model of the proposed device}} --- As depicted in Fig. \ref{f1}(a),
the bottom layer of the proposed device is a 2D semiconductor
with Rashba spin-orbit coupling (SOC), and two superconducting bulks
cover the left and right side of the 2DEG, respectively.
Besides, a metal layer connects these two superconducting bulks
through their top surfaces. The corresponding Hamiltonian has the form of:
\begin{eqnarray}\label{model1}
 \mathcal{H}_{\mathrm{2DEG}}& =& \sum\nolimits_{\mathbf{i},\mathbf{d},\alpha } { (-t_0\psi _{\mathbf{i} + \mathbf{d},\alpha }^\dag  \psi _{\mathbf{i},\alpha }  + h.c.) - \mu \psi _{\mathbf{i},\alpha }^\dag  \psi _{\mathbf{i},\alpha } } \nonumber \\
&-& \sum\nolimits_{\mathbf{i},\mathbf{d},\alpha ,\beta } { i{U _{R}} \psi _{\mathbf{i} + \mathbf{d},\alpha }^\dag  \hat z \cdot (\vec{\sigma}  \times \mathbf{d})_{\alpha \beta }   \psi _{\mathbf{i},\beta } } ,\nonumber  \\
 \mathcal{H}_{s,\mathrm{bulk}} & = & \sum\nolimits_{\mathbf{i},\mathbf{d},\alpha } { (-t_0\psi _{\mathbf{i} + \mathbf{d},\alpha }^\dag  \psi _{\mathbf{i},\alpha }  + h.c.) - \mu \psi _{\mathbf{i},\alpha }^\dag  \psi _{\mathbf{i},\alpha } } \nonumber \\
& +& \sum\nolimits_{\mathbf{i},\alpha} \Delta e^{i\varphi_s} \psi _{\mathbf{i},\alpha }^{\dagger} \psi _{\mathbf{i},-\alpha }^{\dagger}+h.c. ,\nonumber\\
 \mathcal{H}_{N} & = & \sum\nolimits_{\mathbf{i},\mathbf{d},\alpha } { (-t_0\psi _{\mathbf{i} + \mathbf{d},\alpha }^\dag  \psi _{\mathbf{i},\alpha }  + h.c.) - \mu_n \psi _{\mathbf{i},\alpha }^\dag  \psi _{\mathbf{i},\alpha } }. \nonumber  \\
 \end{eqnarray}

\begin{figure}
\centering
\includegraphics[width=3.25in]{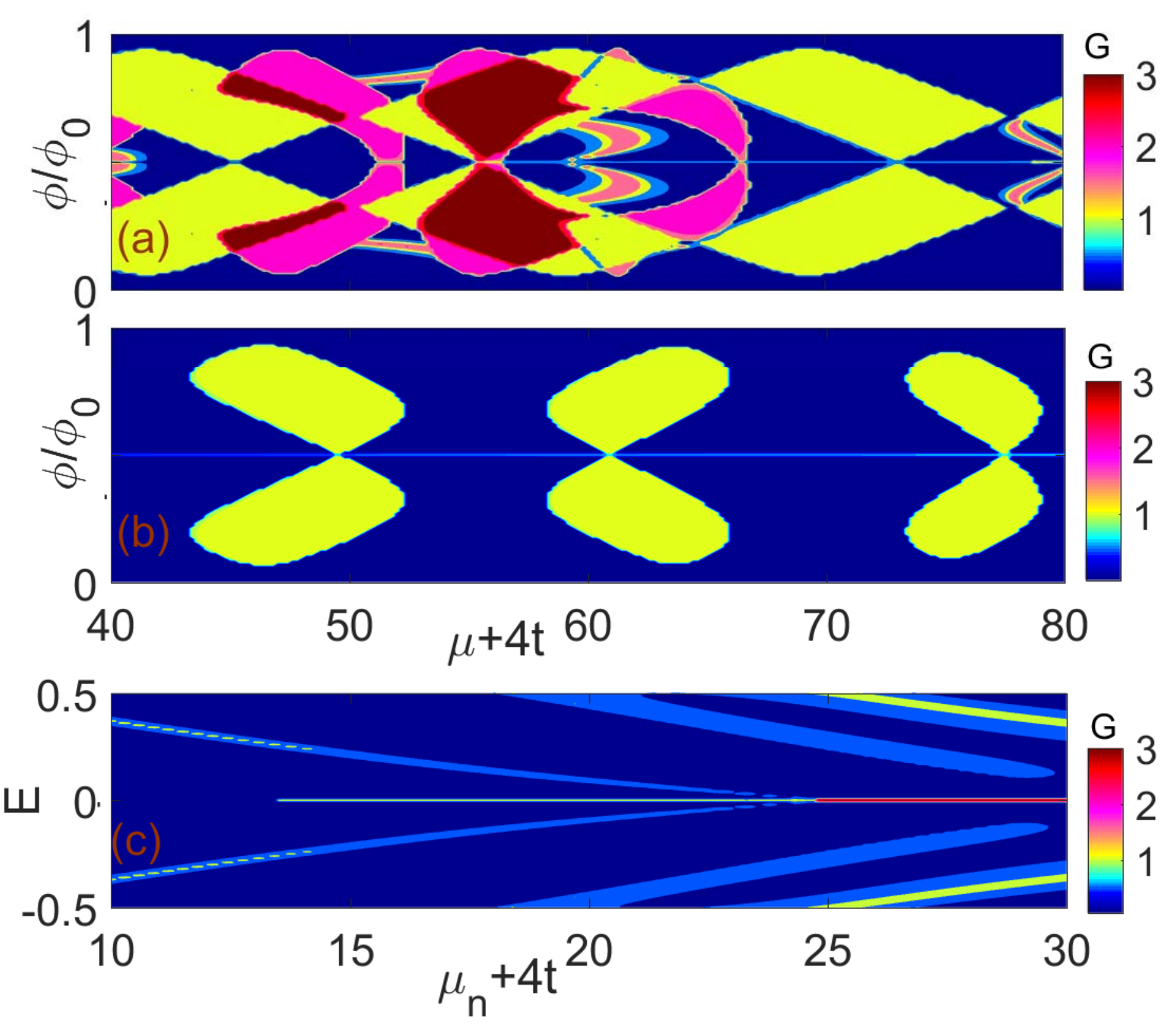}
\caption{
Topological phase diagram of the hollow nanowire structure as functions of the chemical potential in the bottom surface $\mu$ and the flux $\phi$, where the topology is determined by the ZBP. $\Delta = 0.1$meV, phase difference $\varphi=0$, and the sizes $H=5a$, $W=5a$. Rashba SOC is presented only at the bottom surface of the rectangular hollow nanowire.
(a) $\mu_n=-4t+30\Delta$;
(b) $\mu_n=-4t+16\Delta$, and the topologically non-trivial region is significantly changed compared with (a).
(c) The ZBP versus $\mu_n$ and the energy of the incident electron $E$ with $\mu=-4t+46\Delta$. With the decrease of $\mu_n$, the number of MZM pairs reduces from $3$ to $1$, and finally reduces to $0$.
}
\label{f2}
\end{figure}

\noindent Here, $\mathcal{H}_{\mathrm{2DEG}}$ is the 2DEG's Hamiltonian with Rashba SOC which is presented only at the bottom layer. $\mathcal{H}_{s,\mathrm{bulk}}$ with $s=L$ ($R$) is the Hamiltonian of the left (right) superconducting block which lies at $x>W/2$ ($x<-W/2$)
($W$ is the width of the junction). $\mathcal{H}_{N}$ is the Hamiltonian of the normal metal on the top layer, $\mu$ and $\mu_{n}$ are the chemical potentials in 2DEG and the normal metal, respectively, which can be tuned by gate voltages. In addition, the phases of the superconducting order parameter $\Delta e^{i\varphi_s}$ can be different for these two superconducting bulks
(here we set $\varphi_L=\varphi/2$ and $\varphi_R=-\varphi/2$).
Furthermore, $\mathbf{i}$ denotes the lattice site, and $\mathbf{d}$ is the vector connecting the nearest neighbor sites, $\alpha$ and $\beta$ are the spin indices.
$U_{R}$ is the Rashba SOC strength, $\Delta$ is the superconducting pairing amplitude,
and $t_0$ is the hopping amplitude.
Considering that a weak magnetic field $B$ is presented along the $y$-direction, then a flux $\phi=BWH/\phi_0$ threading the hole is induced and $t_0$ is modified into $t_0e^{i\phi}$ for $-W/2<x<W/2$, where $H$ is the height of the junction and the flux quantum $\phi_0=h/2e$.

The proposed device can be simplified into a hollow nanowire
as shown in Fig. \ref{f1}(b).
Let us consider a rectangular hollow nanowire structure with hight $H$ and width $W$,
the surface of the nanowire can be described by two coordinates $(s,y)$,
where $s$ is periodic coordinate along the angular direction of the nanowire,
and $y$ is along the longitudinal direction.
We first consider a simple case with Rashba SOC pointing perpendicular to the surface of the nanowire.
Then the effective Hamiltonian of the hollow nanowire reads $H_s=\frac{(p_s-eA_s)^2+p_y^2}{2m^*}+U_{R}[\sigma_y(p_s-eA_s)-\sigma_xp_y]$.
Due to the antiperiodic boundary condition along the $s$-direction \cite{flux1},
$p_s = (n-1/2)\pi/(H+W)$ where $n\in Z$, and the spectrum is $E_{p_y, n, \pm} = [(n-1/2-\phi/\phi_0)\pi/(H+W)]^2/2m^*+p_y^2/2m^*\pm U_{R}\sqrt{[(n-1/2-\phi/\phi_0)\pi/(H+W)]^2+p_y^2}$.
The modes in such spectrum are always even occupied at $\phi/\phi_0=0$.
While for $\phi/\phi_0 \not= 0$, the magnetic flux plays the role of effective Zeeman field \cite{core1} and removes their degeneracy.
The system becomes topologically non-trivial in the condition that the modes' occupancy number becomes odd in the presence of superconducting pairing term.
As shown in Fig. \ref{f1}(c), the odd occupancy appears at $p_y=0$ when the energy $E$ increases from a black line mode to a red line mode (indicated by blue dashed line).

Remarkably, the magnetic flux also plays the role of effective Zeeman field when the Rashba SOC is only non-zero at the bottom surface [see Fig. \ref{f1}(a)].
In such condition, the energy window for odd occupancy is still quite large as indicated by blue dashed lines shown in Fig. \ref{f1}(d). What's more, in the presence of homogeneous Rashba SOC, if we add two superconductors on the both sides of the hollow nanowire, then the system can be viewed as two parallel superconductor-semiconductor-superconductor Josephson junction. In the presence of magnetic flux, the two MZMs at the bottom and the top surface will hybridize with each other and destroy the topological phase. These troubles can be solved by introducing an inhomogeneous Rashba SOC term, for example, the Rashba SOC is only non-zero at the bottom surface [see Fig. \ref{f1}(a)]. To further study the influence of magnetic flux when Rashba SOC is only non-zero at the bottom surface, we show the energy spectrum at $p_y=0$ versus the flux in Fig. \ref{f1}(e), where the doubly degenerate states split with the increase of the magnetic flux. It indicates that the magnetic flux serves as an effective Zeeman field here. Moreover, such effective Zeeman field is proportional to $(n-1/2-\phi/\phi_0)\pi*U_{R}/(H+W)$, hence the effective Zeeman field is larger in higher subbands, which is quite different from the real Zeeman field inducing same splitting strength for all modes. For higher chemical potential, the topologically non-trivial region will be suppressed and the gap in the topologically non-trivial region tends to collapse due to the large spacing between subbands. To stabilize the topologically non-trivial phase in such case,  larger real Zeeman field, in another word, larger magnetic field is usually required. However, the larger magnetic field would suppress the superconductivity and destroy the TSC. Therefore, magnetic flux playing the role of effective Zeeman field show great advantages in the high chemical potential case. The parameters we adopted in Fig. \ref{f1}(c)-(e) are drawn from experiments [\onlinecite{PJJ2}]: $m^*=0.033 m_e$ and $U_R = 34$meV$\cdot$nm. (Actually, the topologically non-trivial phase appears in a wide range of the size parameters, see Supplemental Material \cite{S1}.)


{\emph {Manipulable topology and stable gap} --- We have shown
that the flux can serve as an effective Zeeman field and display great advantage in the condition of high chemical potential.
Let us further study the topological phase diagram of the hollow nanowire system in the presence of superconducting pairing term.
Here we set the superconducting pairing strength at both
the left and right superconductor bulks as $\Delta=0.1$meV.
Fig. \ref{f2}(a) shows the phase diagram of the system as
a function of flux $\phi$ and chemical potential $\mu$.
Due to the presence of an additional mirror-symmetry about the $y-z$ plane,
the system belongs to class BDI and may supports $N$ pairs of MZM at the ends of the wire. Here we use the zero-bias peak (ZBP) \cite{Jie3,Jie4} to indicate the topology, since the ZBP is $N\frac{2e^2}{h}$ when $N$ pairs of MZMs are presented in the system \cite{S1}. We can see that the topologically non-trivial region is always presented for any chemical potential.
In the condition of high chemical potential, the gap tend to collapse if $\mu=\mu_n$, which is the same as the previous planar Josephson junction. However, such disadvantage can be overcome by manipulating the chemical potential $\mu_n$.
Though further lower the chemical potential in the top surface would reduce the topologically non-trivial region [Fig. \ref{f2}(b)], the bulk gap increases with the decrease of $\mu_n$. Fig. \ref{f2}(c) further shows the ZBP versus $\mu_n$ and the energy of the incident electron $E$ in the condition of $\mu=-4t+46\Delta$. With the decrease of $\mu_n$, the ZBP decreases from $3$ to $1$, and finally decreases to 0. It means that $\mu_n$ provides an efficient way tuning the topology. In addition, the coherence peak of the bulk increases monotonously with the decrease of $\mu_n$. Hence, a large and hard gap can be obtained by tuning $\mu_n$.


\begin{figure}
\centering
\includegraphics[width=3.25in]{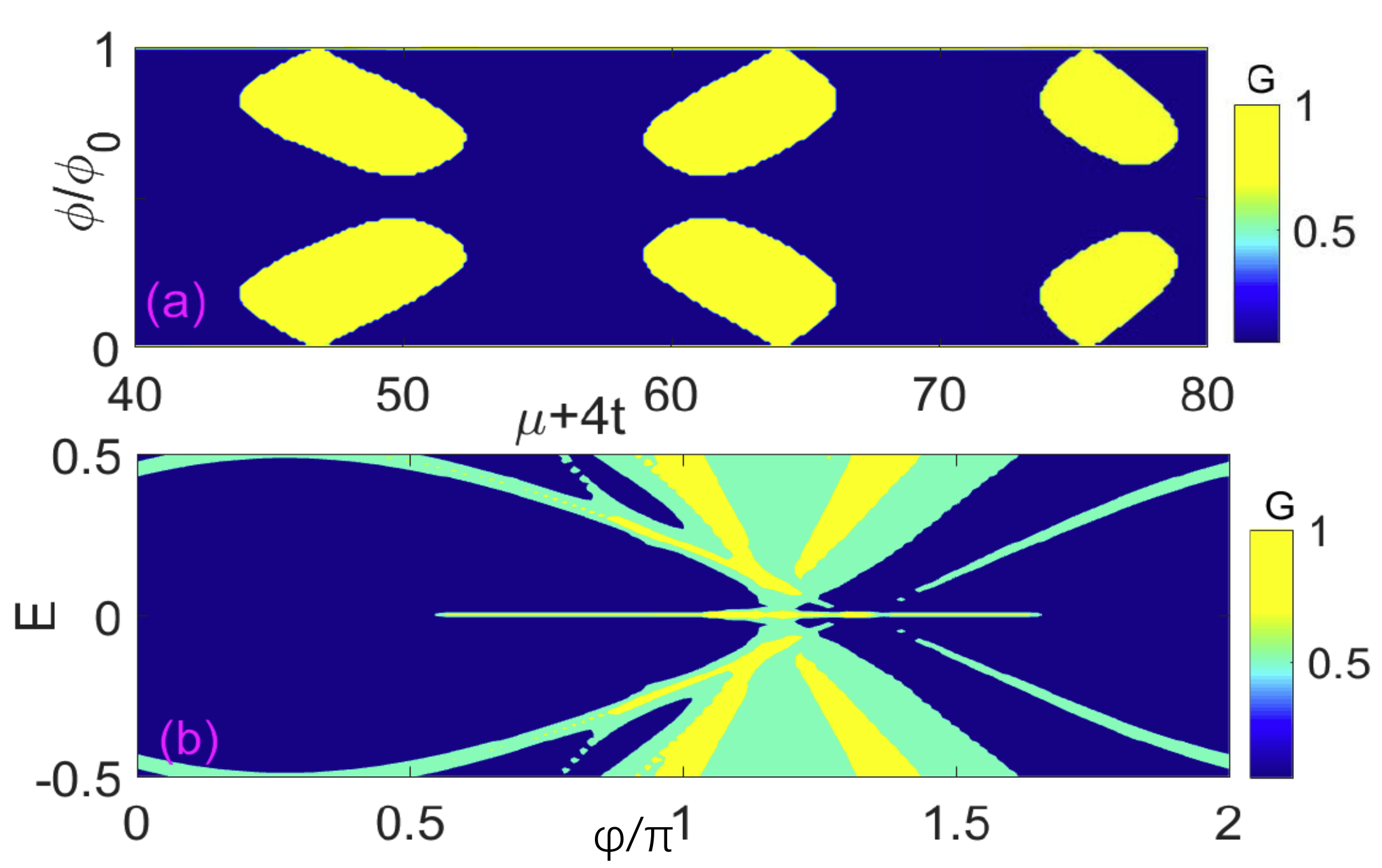}
\caption{(a) ZBP versus the chemical potential in the bottom surface $\mu$ and the flux $\phi$ with $\mu_n=-4t+30\Delta,\varphi=\pi$. The topology is significantly altered with the variation of the phase difference. (b) The ZBP versus phase difference $\varphi$ and the energy of the incident electron $E$ with $\mu=-4t+46\Delta$, $\mu_n = -4t+15\Delta$, and $\phi=0.1\phi_0$.
}
\label{f3}
\end{figure}

Another advantage inherited from the previous planar Josephson junction is
that its topology can be easily tuned by modulating the phase differences $\varphi$.
Fig. \ref{f3}(a) shows the ZBP versus the flux $\phi$ and the chemical potential $\mu$ in the condition of $\varphi=\pi$ and $\mu_n=-4t+16\Delta$.
Comparing with Fig. \ref{f2}(b), the center of the topologically non-trivial region shifts from $\phi=0.5\phi_0$ to $\phi=0$.
The phase $\varphi$ effectively adds a half flux-quantum vortex,
since the flux between the two superconductors will alter
the phase difference between these two superconductors and vice versa.
Therefore, through manipulating the phase difference $\varphi$ between
the two superconducting bulks, we can easily tune the system between
topologically trivial and non-trivial phases.
Such manipulation is clearly exhibited in Fig. \ref{f3}(b),
which shows the ZBP versus the phase difference $\varphi$ and energy of the incident electron $E$
at $\phi=0.1\phi_0$. The ZBP appears at $\varphi=0.5\pi$, indicating that the system transforms from topologically
trivial to topologically non-trivial phase at $\varphi=0.5\pi$.

{\emph {Non-Abelian braiding through cross-shaped junction}} ---
In the previous 2D planar Josephson junction or semiconductor-superconductor nanowire,
the magnetic field can only be presented along one direction,
hence a desired network structure composed of MZMs can only be constructed
through parallel alignment \cite{net1,net2}.
For hollow nanowire Josephson junction, on the contrary, the weak magnetic field
could be presented along both the $x$ and $y$ directions [For the typical parameters adopted in Fig. \ref{f3}(b), the required magnetic field is about 0.01T to generate a flux of $0.1\phi_0$].
Thus, any 2D network structures, such as the honeycomb lattice or square lattice structure composed of MZMs can be easily constructed \cite{HM1,HM2}.

\begin{figure}
\centering
\includegraphics[width=3.25in]{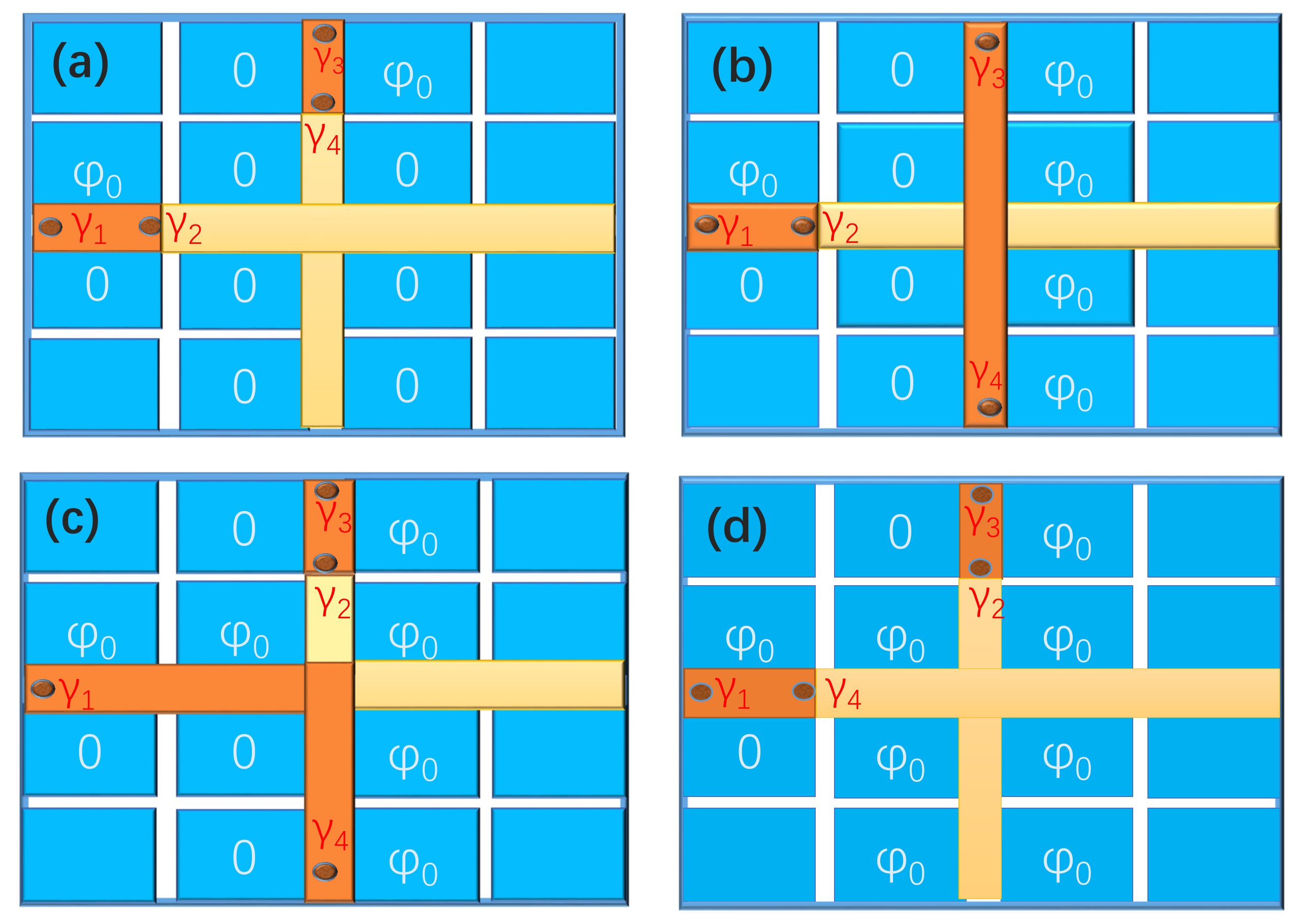}
\caption{Cross-shaped junction for non-Abelian braiding.
(a) Initially, two pairs of MZMs are prepared through modulating the phase differences. The corresponding junction is topologically non-trivial when the phase difference is $\varphi_0$, while topologically trivial when the phase difference is $0$.
(b) Moving MZM $\gamma_4$ to another junction by manipulating the phase differences.
(c) Moving MZM $\gamma_2$ to the original position of $\gamma_4$.
(d) Moving MZM $\gamma_4$ to the original position of $\gamma_2$.}
\label{f4}
\end{figure}

To demonstrate such network structures, we propose a cross-shaped
junction supporting the braiding of MZMs.
As shown in Fig. \ref{f4}, the blue squares are conventional superconductors
on the top surface of the 2DEG, and the cross-shaped junction (covered by metal) is exhibited by the yellow and orange regions. The corresponding junction becomes topologically non-trivial
when the phase difference is $\varphi_0$, while topologically trivial
when the phase difference is $0$ (the flux through the cross section of the junction is fixed).
Fig. \ref{f4}(a) shows the initial condition where two pairs of MZMs are located
at the left side and upper side of the cross-shaped junction, respectively,
by manipulating the phase differences.
Then we firstly move a MZM $\gamma_4$ down to the bottom by tuning the phase
differences of each superconducting bulk as shown in Fig. \ref{f4}(b).
After that, we can move $\gamma_2$ to the original position
of $\gamma_4$ as shown in Fig. \ref{f4}(c).
Finally, as shown in Fig. \ref{f4}(d), $\gamma_4$ is moved to the original position of $\gamma_2$ so that the spatial positions of $\gamma_2$ and $\gamma_4$ are swapped.
We can conveniently braid any pair of MZMs in such structure.
For example, if we want to swap the spatial positions of $\gamma_2$ and $\gamma_3$ in Fig. \ref{f4}(a), we can move both $\gamma_3$ and $\gamma_4$ to the bottom of the junction at first, the following process is similar as the previous process for braiding $\gamma_2$ and $\gamma_4$. The manipulability of such TSC shows great advantages compared with the previous devices.

{\emph {Discussion}} --- We have shown the possibility of constructing TSC by hollow nanowire Josephson junction in the presence of a very weak magnetic field. Such junction provides an additional free parameter manipulating the topology as well as the bulk gap of the system. In addition, the topologically non-trivial region is presented in all energy within the band, and the topology of such TSC can be easily tuned by the phase differences between the superconductors.

In the following, we would like to put forward several advices
which would be helpful for experimental realization.
First, the width of the junction, especially the width of the upper surface should be small, since the large momentum mismatch between the parts with and without Rashba SOC will induce a large bulk gap and supress the additional multi-MZMs.
Thus, a trapezoid structure might be better than the rectangular structure
since it will give rise to larger Thouless energy in the upper surface.
Second, the effective Zeeman energy induced by the flux is typically in the order of $\pi U_R/(H+W)$, which means that a large Rashba SOC would favor the formation of TSC. For conventional semiconductor nanowire or 2DEG, the Rashba SOC strength is in the range of $10 \sim 35$meV$\cdot$nm \cite{RSB1}, which is generally large enough to induce topologically non-trivial phase.
Finally, we would like to point out that  an ideal planar Josephson junction only works well in the limit that the superconducting bulks narrowing down to the nanowire scale \cite{S1}.
A thick superconductor would induce a large self-energy and weaken the effective Zeeman field term.


{\emph{Acknowledgement.}}--- We thank Chui-Zhen Chen, Hua Jiang for fruitful discussion. This work is financially supported by NSFC (Grants No. 11574245, No. 11534001,  and No. 11574007) and NBRPC (Grants No. 2015CB921102, No. 2017YFA0303301).

\end{document}